\documentclass{optica-article}

\journal{opticajournal} % for journals or Optica Open

\articletype{Research Article}

\usepackage{lineno}
\usepackage{siunitx}

\begin{document}

\title{Demonstration of a Mobile Optical Clock Ensemble at Sea}

%\author{
%E. Ahern,\authormark{1} 
%J. W. Allison,\authormark{1}
%C. Billington,\authormark{1}
%N. Bourbeau H\'{e}bert, \authormark{1}
%A. P. Hilton, \authormark{1}
%E. Klantsataya, \authormark{1}
%C. Locke, \authormark{1}
%A. N. Luiten, \authormark{1,*}
%M. Nelligan, \authormark{1}
%R. F. Offer, \authormark{1}
%C. Perrella, \authormark{1}
%S. K. Scholten, \authormark{1}
%B. White, \authormark{1}
%B. M. Sparkes, \authormark{2}
%R. Beard, \authormark{3}
%J. D. Elgin \authormark{4}
%and K. W. Martin \authormark{3}}

\author{
A. P. Hilton,\authormark{1} 
R. F. Offer,\authormark{1} 
E. Klantsataya,\authormark{1} 
S. K. Scholten,\authormark{1} 
N. Bourbeau H\'{e}bert,\authormark{1} 
C. Billington,\authormark{1} 
C. Locke,\authormark{1} 
C. Perrella,\authormark{1} 
M. Nelligan,\authormark{1} 
J. W. Allison,\authormark{1} 
B. White,\authormark{1} 
E. Ahern,\authormark{1} 
K. W. Martin,\authormark{2} 
R. Beard,\authormark{2} 
J. D. Elgin,\authormark{3} 
B. M. Sparkes,\authormark{4} 
and A. N. Luiten\authormark{1,*}}

\authormark{1}Institute for Photonics and Advanced Sensing (IPAS) and School of Physics, Chemistry and Earth Sciences, University of Adelaide, Adelaide, SA 5005, Australia\\
\authormark{2}Blue Halo 1300 Britt Street, Albuquerque, NM  87123 USA\\
\authormark{3}Air Force Research Labs, Space Vehicles Directorate, Quantum Sensing and Timing, 3550 Aberdeen Ave. SE, Kirtland AFB, NM 87117 USA\\
\authormark{4}Defence Science and Technology Group, Edinburgh SA 5111 Australia

%\authormark{1}Institute for Photonics and Advanced Sensing (IPAS) and School of Physics, Chemistry and Earth Sciences, University of Adelaide, Adelaide, SA 5005, Australia\\
%\authormark{2}Defence Science and Technology Group, Edinburgh SA 5111 Australia\\
%\authormark{3}Blue Halo 1300 Britt Street, Albuquerque, NM  87123 USA\\
%\authormark{4}Air Force Research Labs, Space Vehicles Directorate, Quantum Sensing and Timing, 3550 Aberdeen Ave. SE, Kirtland AFB, NM 87117 USA

\email{\authormark{*}email: Andre.Luiten@Adelaide.edu.au} %% email address is required; see note below about the corresponding author designation

% use {asbstract*} to suppress the copyright line. Copyright information will be added in production

\begin{abstract*} 

Atomic clocks are at the leading edge of accuracy and precision and are essential for synchronization of distributed critical infrastructure, position, navigation and timing, and scientific applications.
There has been a breakthrough in the performance of atomic clocks with the shift from microwave to optical frequency transitions. However, this performance increase has come at the cost of size, complexity and fragility, which has confined optical clocks to laboratories. 
Here we report on a recent international collaboration where three emerging optical clocks, each based on different operating principles, were trialled at sea. Over weeks of unsupervised naval operation, these clocks demonstrated exceptional reliability and provided frequency stability outputs in optical, microwave and radio-frequency domains. The performance of all three devices was orders of magnitude superior to existing best-in-class commercial solutions over short and medium timescales, marking a significant step toward deploying optical clocks in real-world environments.

\end{abstract*}

%%%%%%%%%%%%%%%%%%%%%%%%%%  Body  %%%%%%%%%%%%%%%%%%%%%%%%%%
\section{Introduction}
Atomic clocks have been at the leading-edge of accurate and precise measurements since their invention in 1955 \cite{ESSEN1955}.
Following the International Bureau of Weights and Measures' (BIPM) re-definition of the second in 1967 to depend on the cesium ground state transition frequency \cite{Resolution1_1967} atomic clocks have been cemented at the heart of the International System of Units (SI).
The outstanding performance and lower systematic uncertainty associated with optical frequency atomic clocks has led to the expectation that the SI will soon move to a definition for the second based on an optical frequency transition \cite{Resolution5_2022}.  

Since their invention, atomic clocks have increasingly been incorporated into infrastructure that is critical to our daily lives.
The most notable and ubiquitous example is their use in the satellite constellations that form the various Global Navigation Satellite Systems (GNSSs). 
In addition to allowing precise navigation across the globe, atomic clocks are key in providing timing synchronisation across spatially-distributed networks including: telecommunications, distributed computing, electrical power grid synchronisation, relativistic geodesy and very-long baseline radio astronomy applications \cite{ Tavella:20, Schuldt:21 , Boldbaatar:23, RIKEN2016, Rochat:12 }. 
However, the benefits that have accrued from easy and universal availability of GNSS timing signals has perhaps led to an over-reliance in critical infrastructure. 
Furthermore, future high-speed communication and data processing applications will require local timing references with better stability than current microwave clocks can offer \cite{Marin-Palomo:20,Hillerkuss:12}.
To overcome the vulnerabilities and performance shortcomings of GNSS-based timing dissemination on these dependent systems, there is a clear need to develop GNSS-independent, high-performance mobile atomic clocks that can deliver high-quality local timing signals in uncontrolled real-world environments\cite{ Petrov:16, Hasan:18}.

Atomic clocks typically use an oscillator to excite a narrow transition in an ensemble of unperturbed atoms. A signal is derived from this interaction which relates to the frequency difference between the oscillator and the atomic frequency and this is then used to lock the oscillator to the transition frequency.  
Historically, atomic clocks have used a microwave oscillator to drive atomic transitions in the electronic ground state of Cs or Rb atoms \cite{Hellwig1975}. 
However, recent laboratory efforts have focused on using optical transitions with transition frequencies of the order of hundreds of terahertz \cite{Ludlow2015Review}. This new generation of optical atomic clocks has demonstrated a massive leap in frequency stability compared to commercially-available microwave atomic clocks \cite{Ushijima2015, Huntemann2016, Gao2018, McGrew2018, Bothwell2019, Hobson2020}, in part linked to their inherently higher quality-factor (Q-factor).    

However, the optical domain also brings an increase in complexity, and so with a few notable exceptions, optical clocks are large, lab-based devices. 
Recently, optical lattice clocks \cite{LENS2014, SOC3_2018, NTSC2020,  RIKEN2021, Birmingham2022} and ion clocks \cite{Wuhan2017, RAS2022, Opticlock2021} have been shrunk to the scale of a cubic metre, allowing their transport between locations and rapid (circa days) return to operation. 
Vehicle transportable systems have also been demonstrated, such as Physikalisch-Technische Bundesanstalt's (PTB's) transportable strontium lattice clock\cite{PTB2018}, and the Chinese Academy of Sciences' calcium ion clock \cite{Wuhan2020}. In both cases these occupy an air-conditioned trailer and are operable relatively quickly (circa days) after transport.   

Though these devices offer exciting opportunities for the frequency metrology community, usefulness in broader applications in real-world environments requires a truly mobile timing reference that is fully automated and in operation even while in motion. A number of microwave clocks are capable of operating in this regime, with the most commercially successful of these being the Chip Scale Atomic Clock (CSAC) \cite{CSAC} and Microchip Cesium Beam Clock (5071A) \cite{5071A}.  
The first nascent steps for mobile optical clock technology have been taken with a molecular iodine vapor cell, which operated during a six minute flight on a sounding rocket \cite{JOKARUS2019}.

Here we report on a significant step forward in the development of truly mobile optical atomic clocks.  
We have  carried out a marine field trial of an ensemble of three vapour-based optical clocks, each operating on a fundamentally different architecture. This  architectural difference leads to significant variations in environmental sensitivity of each clock strengthening the confidence in our estimation of the long-term frequency stability of the individual clocks. Further, the presence of three simultaneously operating high-performance optical clocks allows us to calculate the short-term performance of each individual optical clock, and thus verify, for the first time, their superiority over commercial microwave counterparts in  deployed and uncontrolled environments. 

The clocks were air-freighted from their respective institutions (two from the University of Adelaide (UofA), Australia, and the third from the Air Force Research Laboratory (AFRL), USA), installed on a ship (HMNZS Aotearoa), and operated autonomously at sea for three weeks. 
We note that along with the three portable optical atomic clocks reported here, a fourth clock technology based on molecular iodine was also demonstrated during the challenge \cite{Roslund2024}.
As we show below, the performance and robustness of the three clocks reported here, along with the good performance of the Vector Atomic technology, emphasizes that a range of robust and compact optical atomic clocks are now ready for real-world deployed applications in applied metrology.

\section{Portable Optical Clocks}

\begin{figure}[htbp]
    \centering
    \includegraphics{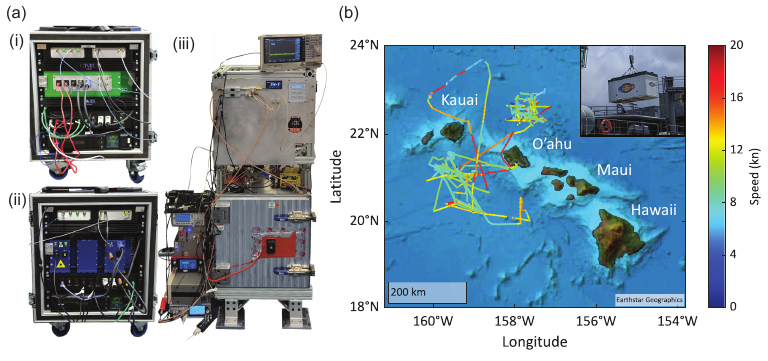}
    \caption{(a) The mobile clock systems: (i) the UofA Yb clock, (ii) the UofA Rb clock, and (iii) the AFRL Optical Rubidium Atomic Frequency Standard (ORAFS). (b) The route taken by the HMNZS Aotearoa during one of the field trials, with an inset showing loading of the Office of Naval Research (ONR) SeaBox onto the deck of the ship via the ship's crane.}
    \label{fig:Trial_Overview}
\end{figure}

The three portable optical clocks that formed the mobile ensemble are pictured in Fig.\,\ref{fig:Trial_Overview}(a). 
Each system contains the key components of an optical atomic clock: a physics package that contains an atomic sample with a high signal-to-noise atomic feature; a photonics package that contains the laser source(s) and photonics to generate a tailored optical interrogation signal along with the digital control that locks the laser frequency to the atomic transition; and an optical frequency comb to down-convert that locked laser signal into radio-frequency (RF) and microwave signal outputs.
All three clocks described here use a thermal vapor cell at the heart of their physics package. 
This avoids the complexity of laser cooling or trapping albeit at the potential expense of a higher environmental sensitivity.  
In each clock the optical transition has a modest linewidth of order \SI{100}{\kHz}, resulting in a sufficiently large Q-factor for high-performance operation, but also allowing high-speed feedback (circa \SI{100}{\kHz})  for active suppression of fast laser noise and conferring high reliability against external vibration and shock during operation. 
 
Two of the clocks were developed by UofA.  
The first of these, the ``Ytterbium Vapour Cell Clock'' (UofA Yb) is based on modulation transfer spectroscopy of the \SI{556}{\nm} inter-combination line in neutral ytterbium vapor \cite{Hilton2024}.  
The second, the ``Rubidium Two-Photon Clock'' (UofA Rb) makes use of a two photon transition in a $^{87}$Rb vapor excited via two counter-propagating lasers at \SI{780}{\nm} and \SI{776}{\nm}  \cite{Perrella2019, Emily2024, Bluebell2024}.
The third clock, the ``Optical Rubidium Atomic Frequency Standard'' (ORAFS) was developed at AFRL and is based on the same two photon transition in $^{87}$Rb vapor, but is excited by a single laser at \SI{778}{\nm}\cite{Martin2018, Lemke2022}. 
Further details of the individual clock technologies can be found in Appendix~\ref{app:Yb},\ref{app:ORAFS} and \ref{app:Rb}.

\section{Maritime Demonstration}
The marine trial of the clock ensemble was undertaken as part of The Technical Cooperation Program (TTCP) Alternative Position, Navigation and Timing Challenge, which was managed by the Office of Naval Research.
This afforded the rare opportunity to test cutting-edge alternative position, navigation and timing technologies in a relevant operational environment. 
During the challenge the clocks were installed within a heavily modified shipping container known as the SeaBox along side other state-of-the-art quantum and precision Position, Navigation and Timing (PNT) experiments from a variety of nations.
The SeaBox was then loaded onto the deck of the HMNZS Aotearoa whilst she actively participated in The Rim of the Pacific Exercise (RIMPAC-2022) in July-August 2022, an international maritime exercise based around Hawaii.

In total, the clocks operated at sea for a period of 21 days covering over 80 nautical miles (150 kilometers).  
This was split over three trips, with the course followed during one of these shown in Fig.\,\ref{fig:Trial_Overview} (b).
During the clocks' time at sea the ship carried out various high-speed manoeuvres and other operations resulting in the clocks experiencing a wide range of acceleration, temperature, humidity, and vibration conditions. 
Accelerations of up to \SI{\pm5}{\m\per\s^2} in a \SI{7}{\Hz} measurement bandwidth in all three axes were recorded by off-the-shelf inertial sensors embedded within the ensemble. 
The clocks were also subjected to significant temperature variations, with daily temperature swings within the thermally isolated clock housings of \SI{3}{\K} at a rate of up to \SI{6}{\K} per hour while the ship was underway, with more rapid temperature swings of up to \SI{10}{\K} recorded whilst the SeaBox was being accessed frequently before the ship left port.
Additionally, the location of the SeaBox on the deck of the ship led to exposure to significant sea spray in adverse weather conditions.

\section{Clock Frequency Measurements}

\begin{figure}[htbp]
    \centering
    \includegraphics{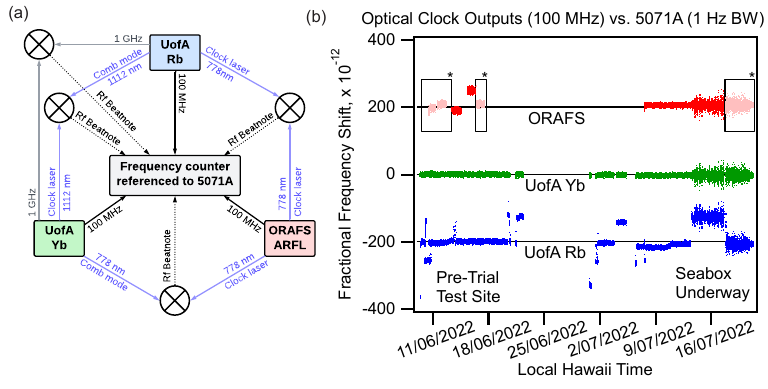}
    \caption{(a) Schematic of the comparison system used to extract the RF and optical frequency stability of each clock, with RF clock outputs (black), microwave outputs (gray), optical outputs (purple) and down-mixed beat notes (dashed).
    (b) Frequency vs time measurement of the \SI{100}{MHz} RF outputs of the three clocks directly measured by the frequency counter during different stages of the trial.
    The short term noise is limited by the 5071A counter reference in all cases.
    Note: The ORAFS data has been processed to account for changes in comb mode index.
    The remaining step changes in frequency are due to the testing of two different locking states, with a small (\num{2e-9} level) difference in frequency offset (denoted by the white box overlays).}
    \label{fig:Frequency_Measurements}
\end{figure}

To assess the performance of the clocks while at sea a versatile frequency comparison system was also installed within the SeaBox.  
The frequency comparison system was designed to allow the frequency stability of each clock within the ensemble to be measured in both the RF and optical domains, with the measurement architecture shown in Fig.\,\ref{fig:Frequency_Measurements} (a).
At the heart of the system is a zero-dead time K+K Messtechnik FXE frequency counter, referenced to a commercial Microchip 5071A microwave cesium beam clock.
Since the performance of each of our optical clocks is intended to exceed that of the 5071A, a simple measurement of the clock output frequencies using this 5071A referenced system would be limited by the noise and drift performance of the 5071A and counter system rather than indicative of the performance of the optical clocks themselves.
However, as we show below, the simultaneous measurement of three high-quality optical clocks allows the use of the three-cornered hat technique \cite{Gray1974}, which in turn allows estimation of the frequency stability of each individual clock without limitation by the 5071A reference.

In total, seven separate frequency measurements were logged simultaneously throughout the duration of the trial as shown in Fig.\,\ref{fig:Frequency_Measurements}(a).
These frequencies relate to frequency differences between the multiple RF and optical frequencies generated by each of the three clocks. 
The in-built optical frequency comb within each clock generated stable RF and microwave outputs at \SI{100}{\MHz} and \SI{1}{\GHz} respectively that were coherently related to the optical output frequencies of the clock. 
Our approach was to directly count the \SI{100}{\MHz} outputs of the three optical clocks, each of which are measurements against the common frequency reference provided by the commercial Cs clock. Although this measurement precision is limited by the performance of the Cs clock, it is possible to eliminate the Cs clock noise after the fact by taking pairwise differences between these measurements. 
In addition, we record the frequency difference between the \SI{1}{\GHz} outputs of the UofA Yb and UofA Rb clocks by mixing these microwave signals down to the RF domain (a direct measurement here would not have sufficient frequency resolution and would have been limited by the counter) and counting the resulting \SI{10}{\kilo\Hz}-range signal.
Finally, we measured the relative frequency fluctuations between the optical frequency outputs of the three clocks. For the two Rb clocks we produced a difference frequency signal between their optical outputs by combining them on a photodetector - the RF output of the photodetector was recorded using the frequency counter. However, given the large difference in output frequencies of the Yb and Rb-based clocks, this direct approach was not possible.  Instead, we used the  integrated optical frequency comb in each of the UofA clocks to coherently synthesize a second optical output that was in the frequency environs of the output of the other clock, as shown in Fig.\,\ref{fig:Frequency_Measurements}(a).  We then measured the frequency difference between the synthesized signal and the second clock by combining them on a photodiode.
More details on the comparison system can be found in Appendix~\ref{app:comp}.

\begin{figure}[htbp]
    \centering
    \includegraphics{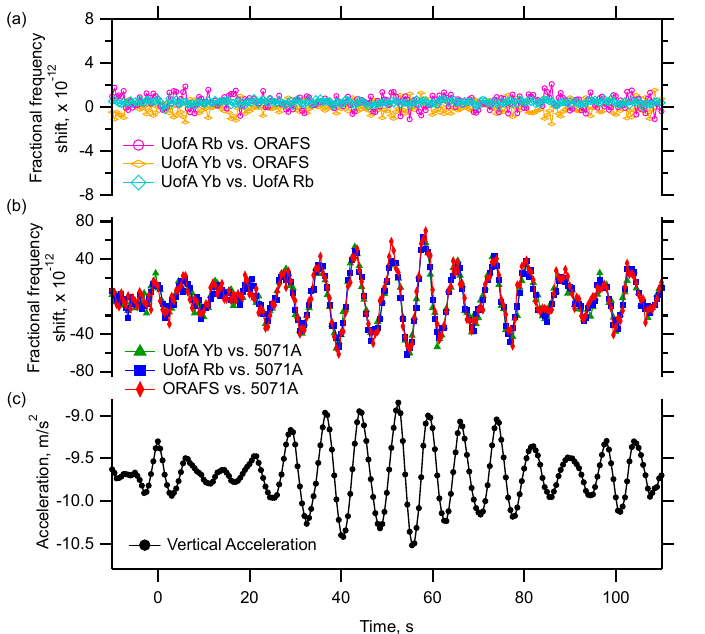}
    \caption{(a) A two minute section of the fractional frequency shift of the optical beats between the three clocks measured by the frequency counter referenced to the 5071A.
    (b) The fractional frequency shifts of the \SI{100}{\MHz} outputs of each optical clock measured directly by the frequency counter referenced to the 5071A. 
    (c) Vertical acceleration measured by a sensor within the clock. 
    }
    \label{fig:Acceleration_Graph}
\end{figure}

The operation and resilience of the clocks is demonstrated in Fig.\,\ref{fig:Frequency_Measurements}(b), where the direct measurements of the \SI{100}{\MHz} output of the clocks is shown as a fractional frequency shift over the total event duration of thirty days.
The noise on short timescales throughout this plot is entirely limited by the 5071A used to reference the frequency counter and is common mode across all three raw clock measurements.
The plot measures clock frequencies during six different phases: a first period of $\sim 10$ days on land within the offices of an off-base contractor facility that provided a baseline measurement of clock performances in a well controlled, air-conditioned environment following transport to Hawaii, a nine day period in which the clock units and other attending quantum technologies were repackaged into provided racks inside the Office of Naval Research's SeaBox at Joint Base Pearl Harbor-Hickam (during which there is no data), a five day period in which the SeaBox was operated on land before it was loaded onto HMNZS Aotearoa, and an eight day operational period of the clocks while in harbour.  
The data after 13/07/22 shows the clock performance while the ship was underway on the route shown in Fig.\,\ref{fig:Trial_Overview}(b).  
This was divided into two four day periods interrupted by an intentional removal of power that required the clocks to shutdown autonomously. 
The AFRL team used on-board personnel to restart the clock while the UofA systems were designed to operate autonomously and automatically recover from a loss-of-power event.
All clocks successfully re-started after this event and remained operational until they were intentionally deactivated once the ship returned to port.  

Considering the measurements in Fig.\,\ref{fig:Frequency_Measurements}(b) as a whole, the Yb clock was highly stable over the entire campaign. Averaged over a \SI{1000}{\s} window, the total fractional frequency deviation was less than \num{6e-12} throughout.
We note the presence of a handful of step changes in the frequency of the UofA Rb and ORAFS over the same period, with the output fractional frequency stepping by \num{6e-11} and \num{2.5e-9}, respectively. 
In the case of UofA Rb clock these step-like features arise after a frequency unlocking event - the mechanism by which this leads to steps is described in more detail below but relates to the means by which the \SI{780}{\nano\meter} laser is frequency stabilized to the modes of the incorporated frequency comb. This is being resolved on the next iteration of the clock.
The step-like features on the ORAFS clock output are a result of changes to the interrogating optical power, again leading to changes in light shift of the two-photon transition.

Also of note in Fig.\,\ref{fig:Frequency_Measurements}(b) is an increased short-timescale noise while the ship was underway (after 13/07/2022). 
As with the rest of the data this is entirely limited by the 5071A and is common between all three measurements.
This is demonstrated in Fig.\,\ref{fig:Acceleration_Graph}, where a two minute section of the frequency of the nominal \SI{100}{\MHz} clock outputs is plotted along with instantaneous vertical acceleration data recorded by sensors embedded within the clocks. 
A strongly correlated \SI{8}{\hertz} modulation is seen in all these signals.
In contrast, the synchronous direct \SI{1}{\GHz} comparison of the UofA clocks shows no evidence of acceleration sensitivity at this level indicating that the increased short time scale noise at sea is associated with the acceleration sensitivity of the 5071A. 

\section{Individual Clock Stabilities}

\begin{figure}[p]
    \centering
    \includegraphics{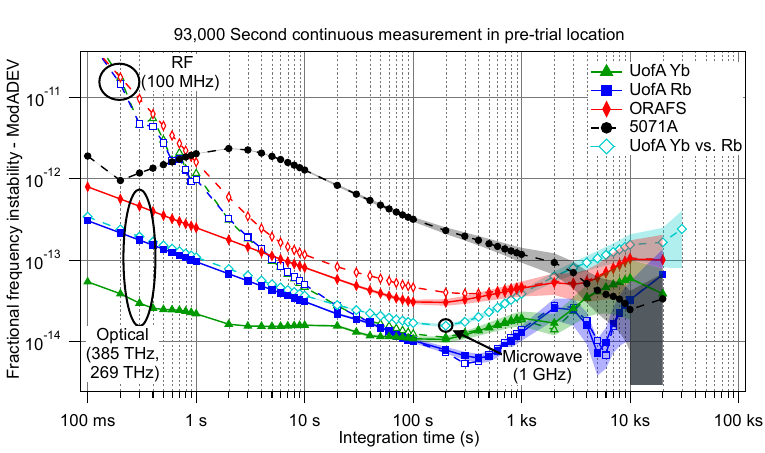}
    \caption{Synthesised individual clock Modified Allan Deviations (ModADEVs) calculated for the three direct optical comparisons (solid lines) and synthesised ModADEVs from \SI{100}{\MHz} measurements (dashed lines) while the clocks were on shore. 5071A counter reference is shown in black dashed line with closed circles. 
    Use of the three-cornered hat method allows extraction of the performance of each individual clock, without noise or drift contributions from the 5071A counter reference. 
    As discussed in text, the \SI{100}{\MHz} measurements are limited by counter noise, while the direct \SI{1}{\GHz} UofA Yb and UofA Rb clock comparison, as well as the three direct inter-clock optical comparisons, are not limited by this noise source.}
    \label{fig:PDR_ADEV}
\end{figure}

\begin{figure}[p]
    \centering
    \includegraphics{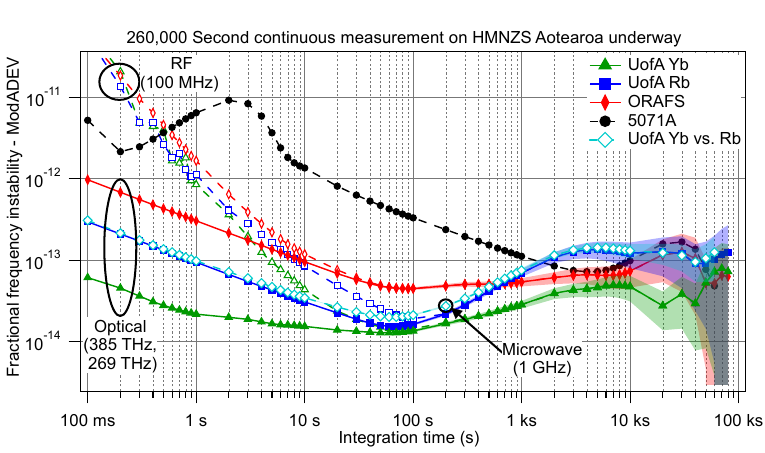}
    \caption{Synthesised individual clock Modified Allan Deviations (ModADEVs) calculated for the three direct optical comparisons (solid lines) and synthesised ModADEVs from \SI{100}{\MHz} measurements (dashed lines) while the clocks were at sea. 5071A counter reference is shown in black dashed line with closed circles.}
    \label{fig:Underway_ADEV}
\end{figure}

A more detailed examination of the frequency stability of the individual clocks was made by carrying out a comprehensive statistical analysis of the frequency measurements.  
The results of this are shown in Figs.\,\ref{fig:PDR_ADEV} and \,\ref{fig:Underway_ADEV} where we compare the RF, microwave and optical performance of the clocks whilst on shore during initial testing (Fig.\,\ref{fig:PDR_ADEV}), and whilst the HMNZS Aotearoa was underway (Fig.\,\ref{fig:Underway_ADEV}).
We estimate the optical performance of each clock by calculating the modified Allan deviation \cite{Allan1981} (ModADEV) for each comparison and then using a three-cornered hat estimation to extract the individual performances \cite{Gray1974}.
We choose the ModADEV measure because it distinguishes between phase flicker noise and white phase noise components, and has a simple relationship to the accumulated timing error.
The three-cornered hat estimated ModADEV for each clock is plotted in Figs.\,\ref{fig:PDR_ADEV} and \,\ref{fig:Underway_ADEV} (solid lines), as well as the direct ModADEV of the \SI{1}{\GHz} UofA Yb vs UofA Rb comparison (yellow). 

Additional processing is required for the RF domain results since the \SI{100}{\MHz} clock outputs were directly counted, and are therefore limited not only by the 5071A reference as discussed earlier, but also by noise introduced by the counter itself. 
To analyse the RF performance we first take frequency differences between independent pairs of the \SI{100}{\MHz} measurements.  
This removes noise and drift contributions from the 5071A and generates three effective comparisons between the \SI{100}{\MHz} clock outputs.
We then extract the RF ModADEV (dashed lines on Figs.\,\ref{fig:PDR_ADEV} and \,\ref{fig:Underway_ADEV}) of each individual clock by using three cornered hat estimation as before.  
As the Cs 5071A is significantly noisier than the other clocks in the comparison we can estimate its performance from the noise in the direct frequency measurements.  This is shown on Figs.\,\ref{fig:PDR_ADEV} and \,\ref{fig:Underway_ADEV} in black.

The integrated frequency combs should faithfully transfer the clocks' optical frequency stability into its RF outputs and hence one would expect the RF and optical ModADEVs to agree exactly.
Whilst true for times longer than \SI{100}{s}, the RF measurements show elevated noise at times shorter than this. 
This is due to counter noise impacting the \SI{100}{\MHz} measurements, which is uncorrelated between counter channels and therefore cannot be removed as common-mode noise.
We confirm this by comparing the UofA Yb vs UofA Rb \SI{1}{GHz} comparison measurement to the UofA Yb and UofA Rb optical performances. 
The higher carrier frequency of these comparisons lowers the contribution of the counter noise to the resulting measurement such that it is no longer setting the measurement limit. 
The \SI{1}{\GHz} measurement shows a fractional frequency stability that is identical to the sum of the optical frequency instabilities of the two UofA clocks for integration times greater than \SI{100}{\milli\s}. 
This demonstrates that the integrated frequency combs are indeed capable of fully transferring the optical stability of the clocks into the RF domain and that the additional instability seen in the \SI{100}{\MHz} RF measurements is associated with the frequency measurement system rather than the clocks themselves. 

All three optical clocks within the ensemble show a significant improvement in performance over the 5071A for integration times shorter than \SI{1000}{\s}.  
For example, the UofA Yb clock shows a stability in the \num{e-14} domain from \SI{100}{\ms} to \SI{80}{\kilo\s} of integration time.
This improvement in performance by over an order of magnitude will be key for future applications such as high-speed communication and data processing. 
Comparing Figs.\,\ref{fig:PDR_ADEV} and \,\ref{fig:Underway_ADEV}, the short-term performance (integration times below \SI{100}{\s}) of each clock at sea was similar to that exhibited in the on-shore test. 
This demonstrates the robustness of each clock and is a clear indicator that these vapor-based optical clocks, including their integrated fibre-based frequency combs, are ready for real-world applications. 
For measurement times beyond \SI{100}{\s}, each clock showed a decreased frequency stability in the moving environment when compared to the stationary tests.
It is difficult to disentangle the effects of environmental influences on individual clocks, as well as the same influences on the inter-clock measurement apparatus.
We note that despite this, these results represent a step-change improvement for portable timing solutions. 
Further, the environmental sensitivity of these prototype systems is not an intrinsic effect associated with the atomic transitions themselves, but instead is associated with the prototype packaging, which will be addressed in further engineering models.

\section{Discussion}

\begin{figure}[htbp]
    \centering
    \includegraphics{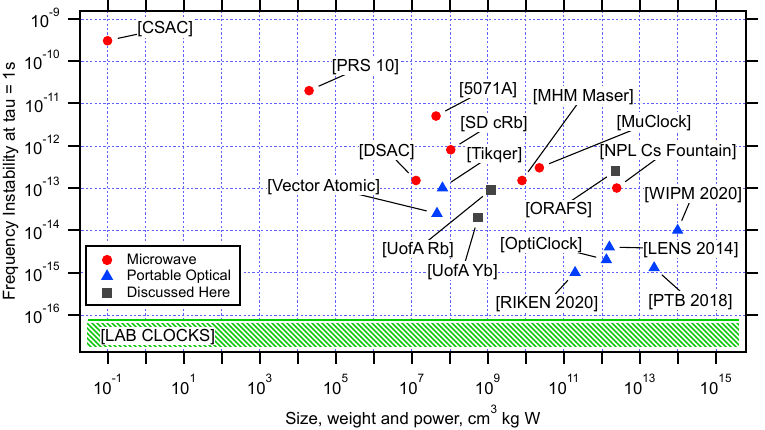}
    \caption{Fractional frequency stability at 1s vs Size-Weight-and-Power of UofA Yb, UofA Rb and ORAFS, compared to a limited selection of commercially available and emerging microwave (red) and optical (blue) portable atomic clocks. 
    CSAC: Microchip Chip scale atomic clock; PRS 10: Stanford Research Systems Low phase noise Rb oscillator; 5071A: Microchip 5071A Primary frequency standard; SD cRb: SpectraDynamics cold Rubidium microwave atomic clock; MuClock: Muquans MuClock; MHM Maser: Microchip MHM-2020 Active hydrogen maser; DSAC: Deep space atomic clock \cite{DSAC2021}; NPL Cs: National Physics Laboratory Cs Fountain \cite{NPLCsFountain}; VA: Vector Atomic Evergreen-30 \cite{Roslund2024}; OptiClock: Yb ion clock \cite{Opticlock2021}; WIPM 2020: Ca ion clock \cite{Wuhan2020}; LENS: Sr Lattice clock \cite{LENS2014}; RIKEN: Sr Lattice clock \cite{RIKEN2020}; PTB: Sr Lattice clock \cite{PTB2018}.}
    \label{fig:SWAP_Plot}
\end{figure}

The three optical clock technologies represented here balance a necessary compromise between the extremely high stability offered by the world's best lab-based optical clocks, against the robustness and practical considerations required to make a portable, deployable device. 
This trade-off is illustrated in Fig.\,\ref{fig:SWAP_Plot} where we compare a range of both microwave and optical clocks as a function of their frequency instability at \SI{1}{\s} and a measure of their combined size-weight-and-power (SWaP) metric in \SI{}{\centi\meter\cubed\kilo\gram\watt}.
The clocks range from compact commercial microwave clocks in the top left, to ultra-high performance transportable optical lattice clocks in the bottom right. 
Previously demonstrated transportable optical clocks are shown in blue although we note that the SWaP remains prohibitively high for deployed applications.

\begin{table}[htbp]
\centering
\caption{\bf Size, weight and power of the three deployed optical atomic clocks}
\small
\begin{tabular}{ccccccc}
\hline
    \bf System & \bf Size (Rack Units) & \bf Size (L) & \bf Weight (kg) & \bf Power (W) \\
\hline
    UofA Yb & 7 & 67 & 40 & 210 \\
    UofA Rb & 7 & 81 & 45 & 340 \\
    AFRL ORAFS & 46 & 760 & 310 & 1300 \\
    %5071A & 3 & 30 & 30 & 50 \\
\hline
\end{tabular}
\label{tab:SWaP_comparisons}
\end{table}

The atomic-vapour-based optical clocks demonstrated here, as well as the other optical clock technology demonstrated in the trial \cite{Roslund2024}, have made a breakthrough reduction in SWaP whilst still making significant improvements in performance over currently available and emerging microwave clocks. 
The SWaP breakdown for UofA Yb, UofA Rb and ORAFS is shown in Table~\ref{tab:SWaP_comparisons}.
The SWaP of these clocks is now such that they can be installed by a single person within existing infrastructure and be easily powered by a single mains outlet.  
In the case of the two UofA clocks, operation during transport both on-foot and within a moving vehicle whilst within their normal cases as seen in Fig\,\ref{fig:Trial_Overview}(a) has also been demonstrated.
Furthermore, each system has reached a sufficient level of automation that they operate without human intervention.

\section{Conclusion}
The results of a field trial of a mobile optical clock ensemble at sea presented here confirm that vapor-based optical clocks with integrated frequency combs are now on the cusp of real-world applications. 
Future development of optical atomic clocks into off-the-shelf devices will deliver wide-ranging benefits providing reliable PNT for vessels operating under a diverse set of conditions, including undersea or in dense urban environments.  
In the commercial sphere, precision timing is critical for any system that requires synchronisation, where portable optical clocks will provide significant competitive advantage.
The mobile and autonomous optical atomic clocks will also support the scientific community through allowing for better tests of fundamental physics, relativistic geodesy, and synchronisation of distributed sensors such as very-long baseline interferometry arrays amongst other applications.

%%%%%%%%%%%%%%%%%%%%%%% Appendix %%%%%%%%%%%%%%%%%%%%%%%%%
\appendix
\section{Appendix}

\subsection{Ytterbium Cell Atomic Clock - UofA Yb}
\label{app:Yb}

The UofA Yb clock is based on the singly-forbidden $^1\mathrm{S}_0 \rightarrow {^3\mathrm{P}_1}$ transition in neutral ytterbium-174, with optical frequency $\nu_0=\SI[group-separator={,}]{539386600191(118)}{\kHz}$ \cite{Atkinson2019}.
This species and transition was chosen because it was amenable to a simple vapor cell architecture, with a narrow clock transition and a convenient wavelength for excitation with commercially available fiber lasers.  
The particular isotope was chosen as it has the largest relative abundance and a desirable electronic configuration in which the lack of nuclear spin combined with the filled external $\mathrm{S}$ shell results in a single ground state with low magnetic sensitivity.
Further, the lack of any nearby single or two-photon transitions from either the ground or excited state provides a low sensitivity to light-shifts. 

The interrogation of the atomic transition is achieved using modulation transfer spectroscopy (MTS), which provides for a high signal to noise ratio.  
Frequency modulation of a strong pump beam is converted to intensity modulation of a weak, counter-propagating probe beam through the optical pumping of the atomic medium. 
Coherent demodulation of the transmitted probe intensity provides for a sub-Doppler error feature with high contrast and low contamination from the broader Doppler background.
 
The Yb interrogating laser system consists of a seed fiber laser at \SI{1112}{\nm}, which is amplified and frequency doubled to \SI{556}{\nm} using a fiber-coupled waveguide second harmonic generator module (SHG).
The \SI{556}{\nm} light is split into pump and probe arms - the pump arm is frequency-modulated by an acousto-optic modulator (AOM), while the probe arm passes through a voltage-controlled optical attenuator (VOA).
These actuators provide the ability to control the optical power.
The combination of these two actuators provides the ability to control the power of both beams, while the AOM additionally allows for residual amplitude modulation (RAM) suppression.
The pump and probe beams are directed to a hot Yb vapour cell.
All input and output signals are monitored and a pair of field-programmable gate array (FPGA) signal processing units are used to maintain the necessary control loops for power and frequency stabilisation.
The direct clock output is taken from a sample of the \SI{1112}{\nm} light before the SHG unit.  

The output signal of this clock provides an optical frequency reference for a self-referenced optical frequency comb based on the National Institute for Standards and Technology (NIST) design\,\cite{sinclair2015invited}. 
The repetition rate of this frequency comb is steered to ensure that the phase of the nearest optical frequency mode of the comb is locked to the phase of the Yb clock output. 
The pulse train output of the comb is also measured on a high-speed photodetector from which we deliver stable RF frequencies at the base repetition rate of the comb (\SI{200}{\MHz}) as well as its higher harmonics.
We specifically choose the fifth harmonic of the repetition rate at \SI{1}{\GHz} to allow for high sensitivity microwave clock comparisons.
The \SI{200}{\MHz} output is also divided to \SI{100}{\MHz} and \SI{10}{\MHz} using the FPGA unit to provide standard countable RF outputs.

Both the clock and comb control systems are managed by a dedicated mini-PC embedded in the larger clock package which performs oversight tasks such as a cold-start procedure, status logging, error state detection, and re-start protocols. 
The clock system is entirely autonomous, with the transition from a cold start to a fully operational clock achieved in 30 minutes via key-turn operation. 
The Yb clock has run independently for months at a time without human intervention. 
The physical package occupies 7 rack units in a standard 19" rack with a total volume of \SI{67}{\litre}, including clock, comb, oversight and power supply, with a weight of \SI{40}{\kg} and wall-socket power consumption of \SI{210}{\W}.  

\subsection{Optical Rubidium Atomic Frequency Standard - ORAFS}
\label{app:ORAFS}
% Note this was edited by Sarah Scholten to match the other section formats - NEEDS CHECKING BY AFRL ORAFS TEAM AS MANY ASSUMPTIONS HAVE BEEN MADE

%The AFRL ORAFS clock was based upon the design previously described in \cite{Martin2018, Lemke2022} where the control electronics were integrated into an FPGA and placed in vacuum along with the laser and vapor cell physics package.
%The lasers, vapor cell and electronics operate on a 28 volt 3 amp dc power supply and were designed to occupy a volume of 68 liters.
%However, since they were designed for a vacuum environment a vacuum chamber and appropriate vacuum pumps and heat rejection plate were necessary for proper clock operation.
%Additionally, the FPGA providing fast feedback to the laser current to the two photon atomic transition was excessively noisy and we had to revert back to a commercial off the shelf laser driver and locking electronics for the demonstration. 

The AFRL ORAFS clock is based upon the design previously described in \cite{Martin2018, Lemke2022}.
ORAFS uses  counter-propagating \SI{778.1}{\nm} beams to perform Doppler-free excitation of the two-photon $5\mathrm{S}_{1/2}\rightarrow5\mathrm{D}_{5/2}$ transition in $^{87}$Rb. 
A telecommunications fiber laser at \SI{1556.2}{\nm} is amplified using a semiconductor optical amplifier (SOA), and then frequency-doubled via a SHG to generate the \SI{778.1}{\nm} light used to excite the two-photon transition. 
The laser power delivered to the Rb atoms is detected just before the Rb cell and then actively stabilized by feeding back to the current of the SOA, which modulates the gain, and hence the total optical power.  
The interrogating laser is frequency modulated and one observes consequent intensity modulation of a \SI{420}{\nm} fluorescence produced via the $6\mathrm{P}_{3/2}$ decay pathway, whose complex amplitude is indicative of the average detuning of the \SI{778}{\nm} laser signal from exact two-photon resonance. 
Synchronous demodulation of the amplitude of this fluorescence signal delivers a signal proportional to this frequency detuning.  
A commercial-off-the-shelf laser driver and locking electronics were used to lock the frequency of the \SI{1556.2}{\nm} laser to that of the atomic transition.  
 
The other control systems and control electronics were implemented on an FPGA platform and placed in a vacuum along with the laser and vapor cell physics package.
The lasers, vapor cell and electronics occupy a volume of \SI{112}{\litre}, weigh \SI{30}{\kg}, and consume \SI{300}{\watt} of electrical power. 
We note, however, that since this clock was originally designed to operate in a vacuum environment, it was necessary to operate the clock within a vacuum chamber together with appropriate vacuum pumps and a heat rejection plate. 
The total footprint of system including ancillary vacuum apparatus occupies a 46 rack units, weighed \SI{310}{\kg}, and consumed \SI{1300}{\watt} of wall-socket power.
An in-built self-referenced optical frequency comb has its repetition rate actively steered so that one mode of the comb is frequency stabilized to the \SI{1556.2}{\nm} laser output. 
Further, by detecting the fundamental repetition rate of the comb one automatically obtains a highly stable \SI{100}{\MHz} stable microwave output for comparison purposes with other RF sources. 

\subsection{Rubidium Two-Photon Clock - UofA Rb}
\label{app:Rb}

The UofA Rb clock utilises dual-color excitation of the two-photon $5\mathrm{S}_{1/2}\rightarrow5\mathrm{D}_{5/2}$ transition in $^{87}$Rb. 
The excitation light is derived from two telecommunications lasers at \SI{1552}{\nm} and \SI{1560}{\nm} by frequency doubling using two SHGs, with the resulting \SI{776}{\nm} and \SI{780}{\nm} beams launched in opposing directions in the vapor cell to minimize Doppler broadening from the atomic motion.   
The \SI{1560}{\nm} laser is frequency locked to a mode of the integrated optical frequency comb to give the \SI{1560}{\nm} laser a fractional frequency stability in the \num{e-13} range. 
We frequency modulate the \SI{1552}{\nm} laser, and observe synchronous intensity modulation of the \SI{420}{\nm} fluorescence from the $6\mathrm{P}_{3/2}$ decay pathway when the sum frequency of the two lasers is near two-photon resonance. 
This fluorescent light is detected with photomultiplier tubes and then demodulated with the frequency modulation signal.  
The resulting error signal allows us to steer the frequency of the \SI{1552}{\nm} laser so that the sum of the \SI{776}{\nm} and \SI{780}{\nm}  photons corresponds to exact two-photon resonance. 
This two-colour excitation deliver freedom to tune the \SI{780}{\nm} laser so that the first step of the two-photon excitation is nearly co-incident with the $5\mathrm{S}_{1/2} \rightarrow 5\mathrm{P}_{3/2}$ transition in Rb. 
This near-resonant first step gives a large enhancement of the two-photon transition rate for a given input power.
The disadvantage of the two-colour approach is that it introduces an unwanted residual Doppler broadening ($\sim \SI{4}{\MHz}$) to the transition width along with introducing the extra complexity of two lasers. 
Two AOMs are used to stabilize the powers of the two lasers, with the \SI{1552}{\nm} AOM additionally providing frequency modulation of the \SI{1552}{\nm} light. 
The  AOM-based modulation introduces unwanted RAM, which is detected along with the optical powers of the two lasers on photodetectors near to the Rb cell.  
We actively steer the RF drive level of the \SI{1552}{\nm} AOM to suppress the unwanted amplitude modulation.  
The physics package is actively temperature stabilized to maintain constant Rb density within the cell. 
This clock architecture is described in more detail in refs.\,\cite{Perrella2019, Emily2024, Bluebell2024}.

A small portion of the \SI{1552}{\nm} and \SI{1560}{\nm} light is directed through a sum frequency generator (SFG) to generate the stable optical output of the clock at \SI{778}{\nm}.  
This can be used for comparison purposes and was used for the direct optical comparisons with the ORAFS clock, as well as with the UofA Yb comb.
This \SI{778}{\nm} clock output light is then used to stabilize the repetition rate of the dedicated self-referenced optical frequency comb via a phase-locked loop (similarly to that described previously in both the ORAFS and Yb clocks).  
As with the UofA Yb clock, the comb design was based on that from \cite{sinclair2015invited}.  

All stabilisation feedback loops are implemented using software running on FPGA platforms with a master control computer running an automation and oversight script. 
This allowed the clock to cold-start, stabilize all feedback loops, and provided for error handling autonomously. 
This approach also allowed for remote monitoring during some phases of the sea trial. 
The clock system is entirely autonomous, with the transition from a cold start to a fully operational clock achieved in 30 minutes via key-turn operation. As mentioned above, and evident on Fig.\,\ref{fig:Frequency_Measurements}, are step-like features in the optical frequency output of the Rb clock after the autonomous system re-locks the lasers to the transition.  This unwanted feature arises because in the relocking of the \SI{1560}{\nm} laser insufficient control was imposed to ensure that it relocked to the same mode of the frequency comb on each occasion. This discrete change in frequency of the \SI{1560}{\nm} laser resulted in a discrete change in the resulting light shift of the clock transition because of the change in detuning of the \SI{1560}{\nm} laser from the $5\mathrm{S}_{1/2} \rightarrow 5\mathrm{P}_{3/2}$ transition in Rb. In the next iteration of the clock this will be prevented by ensuring that the autonomous system relocks the laser to a defined comb mode.   

The Rb clock has operated independently for months at a time without human intervention or restarts. 
The physical package occupies 7 rack units in a standard 19" rack with a total volume of \SI{81}{\litre}, including clock, comb, oversight and power supply, with a weight of \SI{45}{\kg} and wall-socket power consumption of \SI{340}{\W}. 

\subsection{Clock Comparisons and Analysis}
\label{app:comp}

The phase/frequency counter is a high performance device by K+K \cite{Kramer:01, Kramer:04} that allows simultaneous zero-dead time frequency measurements of 8 inputs in comparison to an external \SI{10}{\MHz} clock input provided by the commercial Microchip Cs beam clock (5071A).
This measurement configuration is shown in Fig.\,\ref{fig:Frequency_Measurements}.
We poll the counter at \SI{100}{\ms} intervals and record the data to the disk of a dedicated rugged computer for later processing.
This allowed uninterrupted measurements over the duration of the exercise even in the event of an individual clock restart.

In addition to the direct measurement of each clock's \SI{100}{\MHz} output, optical and microwave comparisons between the clocks were also recorded. 
For the optical frequency comparison between the UofA Yb and Rb clocks, we undertook the comparison at double the wavelength of the Yb transition (\SI{1112}{\nm}), with this wavelength directly available from the Yb optical clock and accessible as a comb-mode from the UofA Rb clock's optical frequency comb.
For the optical frequency comparison between the two Rb clocks, we made use of a direct optical comparison between the \SI{778}{\nm} optical outputs of the two clocks, without the need of an optical frequency comb.  
The ORAFS to UofA Yb clock comparison was made at \SI{778}{\nm} through a beat note between the ORAFS \SI{778}{\nm} clock output and an optical comb mode from the Yb clock generated by frequency doubling of the stable comb output centred around \SI{1556}{\nm}. 
Although complex, this system ensured that the counter always had available at least one strong difference frequency signal from each of the clock comparisons.  

For the analysis of the optical frequency comparisons, the measured direct difference frequency signals were processed by scaling by the optical frequency and removing the mean frequency to produce fractional frequency data.
This time-series data is then passed through the modified Allan deviation algorithm (ModADEV) \cite{Allan1981} to produce the conventional expression of frequency instability as a function of integration time. 
As we have three pairwise comparisons between three non-identical clocks it is possible to perform a three-cornered-hat analysis \cite{Gray1974} - taking linear combinations of those comparisons to estimate the instability contribution associated with each individual clock.
This technique assumes uncorrelated data, which we expect to be a good assumption over short time scales, however due to the co-located nature of the clocks, and the dramatic environmental variations they all experienced, it is likely that over long time scales this approximation will fail (as would be the case for any co-located clock comparisons).
Nonetheless, it is worth noting that since the three clocks presented here differ in construction, atomic species and in interrogation protocol, this has potentially less susceptibility to systematic errors of this type, since the environmental sensitivities of each system are not inherently matched.

%%%%%%%%%%%%%%%%%%%%%%% Backmatter %%%%%%%%%%%%%%%%%%%%%%%%%
\begin{backmatter}
\bmsection{Funding}
This research was supported by the Australian Government through the Next Generation Technologies Fund (now managed through ASCA).

\bmsection{Acknowledgments}
The data presented here was collected at a TTCP Intelligence Surveillance Target Acquisition and Reconnaissance (ISTAR) group Alternative Position, Navigation and Timing (APNT) event at the Rim of the Pacific (RIMPAC) exercise out of Pearl Harbor, Hawaii in July-August 2022. 
TTCP is a five nation (Australia, Canada, New Zealand, United Kingdom, United States) Defence Science and Technology collaborative research programme. 
We would particularly like to thank the DSTG staff who supported the exercise, including Joe Verringer, Vance Crook, Joanne Harrison, and Ken Grant.

We would like to thank Tommy Willis from the Office of Naval Research for organising the APNT Challenge and providing the SEABOX, and David Collier for managing the day-to-day access and facilities during the demonstration.

We thank the Optofab node of the Australian National Fabrication Facility (ANFF) which utilize Commonwealth and South Australia State Government funding. The authors thank Evan Johnson, Alastair Dowler, and Lijesh Thomas and the rest of the Optofab team for their technical support.

The authors thank Pacific Rim Defense for provide the space, access, and amenities used for the pre-trial benchmarking tests.

\bmsection{Author Contributions}
A.P.H., R.F.O., S.K.S., and E.K. wrote the manuscript.
A.P.H., R.F.O., E.K., and B.W. developed the UofA ytterbium clock. 
S.K.S., C.L., C.P., and E.A. developed the UofA rubidium clock.
N.B.H. and C.L. developed and built the UofA frequency combs.
N.B.H. and C.B. provided software and control solutions for the UofA clocks.
M.N. and J.W.A. provided engineering solutions for the UofA clocks.
K.W.M., R.B., and J.D.E developed the AFRL clock and managed their participation in the trial.
B.M.S. coordinated the field trial.
A.N.L. lead the UofA clocks team and their managed their participation in the trial.

\bmsection{Disclosures}
Approved for public release, distribution is unlimited.  Public Affairs release approval \# AFRL-2024-2933.

The views expressed are those of the authors and do not necessarily reflect the official policy or position of the Department of the Air Force, the Department of Defense, or the U.S. government.

The intellectual property related to the Rubidium Two-Photon Clock is owned by QuantX Labs and covered by patent US 10353270 B2.

\bmsection{Data Availability Statement}
Data underlying the results presented in this paper are not publicly available at this time but may be obtained from the authors upon reasonable request.

\end{backmatter}

%%%%%%%%%%%%%%%%%%%%%%% References %%%%%%%%%%%%%%%%%%%%%%%%%
\bibliography{references}

\begin{thebibliography}{10}
\newcommand{\enquote}[1]{``#1''}

\bibitem{ESSEN1955}
L.~Essen and J.~V.~L. Parry, \enquote{An atomic standard of frequency and time interval: A c{\ae}sium resonator,} {\protect\JournalTitle{Nature}} \textbf{176}, 280--282 (1955).

\bibitem{Resolution1_1967}
\enquote{On the future redefinition of the second,} SI unit of time (second), https://www.bipm.org/en/committees/cg/cgpm/13-1967/resolution-1.

\bibitem{Resolution5_2022}
\enquote{On the future redefinition of the second,} Resolution 5 of the 27th CGPM (2022), https://www.bipm.org/en/cgpm-2022/resolution-5.

\bibitem{Tavella:20}
P.~Tavella and G.~Petit, \enquote{Precise time scales and navigation systems: mutual benefits of timekeeping and positioning,} {\protect\JournalTitle{Satellite Navigation}} \textbf{1}, 1--12 (2020).

\bibitem{Schuldt:21}
T.~Schuldt, M.~Gohlke, M.~Oswald, \emph{et~al.}, \enquote{Optical clock technologies for global navigation satellite systems,} {\protect\JournalTitle{GPS solutions}} \textbf{25}, 1--11 (2021).

\bibitem{Boldbaatar:23}
E.~Boldbaatar, D.~Grant, S.~Choy, \emph{et~al.}, \enquote{Evaluating optical clock performance for gnss positioning,} {\protect\JournalTitle{Sensors}} \textbf{23}, 5998 (2023).

\bibitem{RIKEN2016}
T.~Takano, M.~Takamoto, I.~Ushijima, \emph{et~al.}, \enquote{Geopotential measurements with synchronously linked optical lattice clocks,} {\protect\JournalTitle{Nature Photonics}} \textbf{10}, 662--666 (2016).

\bibitem{Rochat:12}
P.~Rochat, F.~Droz, Q.~Wang, and S.~Froidevaux, \enquote{Atomic clocks and timing systems in global navigation satellite systems,} in \emph{Proceedings of European navigation conference, Gdansk, April,}  (2012), pp. 1--11.

\bibitem{Marin-Palomo:20}
P.~Marin-Palomo, J.~N. Kemal, T.~J. Kippenberg, \emph{et~al.}, \enquote{Performance of chip-scale optical frequency comb generators in coherent wdm communications,} {\protect\JournalTitle{Opt. Express}} \textbf{28}, 12897--12910 (2020).

\bibitem{Hillerkuss:12}
D.~Hillerkuss, R.~Schmogrow, M.~Meyer, \emph{et~al.}, \enquote{Single-laser 32.5\&\#x00a0;tbit/s nyquist wdm transmission,} {\protect\JournalTitle{J. Opt. Commun. Netw.}} \textbf{4}, 715--723 (2012).

\bibitem{Petrov:16}
D.~Petrov, S.~Melnik, and T.~H{\"a}m{\"a}l{\"a}inen, \enquote{Distributed gnss-based time synchronization and applications,} in \emph{2016 8th International Congress on Ultra Modern Telecommunications and Control Systems and Workshops (ICUMT),}  (IEEE, 2016), pp. 130--134.

\bibitem{Hasan:18}
K.~F. Hasan, Y.~Feng, and Y.-C. Tian, \enquote{Gnss time synchronization in vehicular ad-hoc networks: Benefits and feasibility,} {\protect\JournalTitle{IEEE Transactions on Intelligent Transportation Systems}} \textbf{19}, 3915--3924 (2018).

\bibitem{Hellwig1975}
H.~W. Hellwig, \enquote{Atomic frequency standards: A survey,} {\protect\JournalTitle{Proceedings of the IEEE}} \textbf{63}, 212--229 (1975).

\bibitem{Ludlow2015Review}
A.~D. Ludlow, M.~M. Boyd, J.~Ye, \emph{et~al.}, \enquote{Optical atomic clocks,} {\protect\JournalTitle{Rev. Mod. Phys.}} \textbf{87}, 637--701 (2015).

\bibitem{Ushijima2015}
I.~Ushijima, M.~Takamoto, M.~Das, \emph{et~al.}, \enquote{Cryogenic optical lattice clocks,} {\protect\JournalTitle{Nature Photonics}} \textbf{9}, 185--189 (2015).

\bibitem{Huntemann2016}
N.~Huntemann, C.~Sanner, B.~Lipphardt, \emph{et~al.}, \enquote{Single-ion atomic clock with $3\ifmmode\times\else\texttimes\fi{}{10}^{\ensuremath{-}18}$ systematic uncertainty,} {\protect\JournalTitle{Phys. Rev. Lett.}} \textbf{116}, 063001 (2016).

\bibitem{Gao2018}
Q.~Gao, M.~Zhou, C.~Han, \emph{et~al.}, \enquote{Systematic evaluation of a 171yb optical clock by synchronous comparison between two lattice systems,} {\protect\JournalTitle{Scientific Reports}} \textbf{8}, 8022 (2018).

\bibitem{McGrew2018}
W.~F. McGrew, X.~Zhang, R.~J. Fasano, \emph{et~al.}, \enquote{Atomic clock performance enabling geodesy below the centimetre level,} {\protect\JournalTitle{Nature}} \textbf{564}, 87--90 (2018).

\bibitem{Bothwell2019}
T.~Bothwell, D.~Kedar, E.~Oelker, \emph{et~al.}, \enquote{Jila sri optical lattice clock with uncertainty of,} {\protect\JournalTitle{Metrologia}} \textbf{56}, 065004 (2019).

\bibitem{Hobson2020}
R.~Hobson, W.~Bowden, A.~Vianello, \emph{et~al.}, \enquote{A strontium optical lattice clock with 1 × 10-17 uncertainty and measurement of its absolute frequency,} {\protect\JournalTitle{Metrologia}} \textbf{57}, 065026 (2020).

\bibitem{LENS2014}
N.~Poli, M.~Schioppo, S.~Vogt, \emph{et~al.}, \enquote{A transportable strontium optical lattice clock,} {\protect\JournalTitle{Applied Physics B}} \textbf{117}, 1107--1116 (2014).

\bibitem{SOC3_2018}
S.~Origlia, M.~S. Pramod, S.~Schiller, \emph{et~al.}, \enquote{Towards an optical clock for space: Compact, high-performance optical lattice clock based on bosonic atoms,} {\protect\JournalTitle{Phys. Rev. A}} \textbf{98}, 053443 (2018).

\bibitem{NTSC2020}
D.-H. Kong, Z.-H. Wang, F.~Guo, \emph{et~al.}, \enquote{A transportable optical lattice clock at the national time service center*,} {\protect\JournalTitle{Chinese Physics B}} \textbf{29}, 070602 (2020).

\bibitem{RIKEN2021}
N.~Ohmae, M.~Takamoto, Y.~Takahashi, \emph{et~al.}, \enquote{Transportable strontium optical lattice clocks operated outside laboratory at the level of $10^{-18}$ uncertainty,} {\protect\JournalTitle{Nature Photonics}} \textbf{14}, 411--415 (2020).

\bibitem{Birmingham2022}
Y.~B. Kale, A.~Singh, M.~Gellesch, \emph{et~al.}, \enquote{Field deployable atomics package for an optical lattice clock,} {\protect\JournalTitle{Quantum Science and Technology}} \textbf{7}, 045004 (2022).

\bibitem{Wuhan2017}
J.~Cao, P.~Zhang, J.~Shang, \emph{et~al.}, \enquote{A compact, transportable single-ion optical clock with $7.8 \times 10^{-17}$ systematic uncertainty,} {\protect\JournalTitle{Applied Physics B}} \textbf{123}, 112 (2017).

\bibitem{RAS2022}
K.~Khabarova, D.~Kryuchkov, A.~Borisenko, \emph{et~al.}, \enquote{Toward a new generation of compact transportable yb+ optical clocks,} {\protect\JournalTitle{Symmetry}} \textbf{14} (2022).

\bibitem{Opticlock2021}
J.~Stuhler, M.~{Abdel Hafiz}, B.~Arar, \emph{et~al.}, \enquote{Opticlock: Transportable and easy-to-operate optical single-ion clock,} {\protect\JournalTitle{Measurement: Sensors}} \textbf{18}, 100264 (2021).

\bibitem{PTB2018}
J.~Grotti, S.~Koller, S.~Vogt, \emph{et~al.}, \enquote{Geodesy and metrology with a transportable optical clock,} {\protect\JournalTitle{Nature Physics}} \textbf{14}, 437--441 (2018).

\bibitem{Wuhan2020}
Y.~Huang, H.~Zhang, B.~Zhang, \emph{et~al.}, \enquote{Geopotential measurement with a robust, transportable ${\mathrm{ca}}^{+}$ optical clock,} {\protect\JournalTitle{Phys. Rev. A}} \textbf{102}, 050802 (2020).

\bibitem{CSAC}
Microchip, \enquote{{Low-Noise Chip-Scale Atomic Clock - Datasheet},}  (2023).

\bibitem{5071A}
Microchip, \enquote{{5071A Cesium Primary Frequency Standard - Sell Sheet},}  (2023).

\bibitem{JOKARUS2019}
K.~D\"oringshoff, F.~B. Gutsch, V.~Schkolnik, \emph{et~al.}, \enquote{Iodine frequency reference on a sounding rocket,} {\protect\JournalTitle{Phys. Rev. Appl.}} \textbf{11}, 054068 (2019).

\bibitem{Roslund2024}
J.~D. Roslund, A.~Cing\"{o}z, W.~D. Lunden, \emph{et~al.}, \enquote{Optical clocks at sea,} {\protect\JournalTitle{Nature}} \textbf{628}, 736--740 (2024).

\bibitem{Hilton2024}
A.~Hilton, R.~Offer, N.~B. H\'{e}bert, \emph{et~al.}, \enquote{An ytterbium vapor cell clock for deployable applications,} {\protect\JournalTitle{In preparation}}  (2024).

\bibitem{Perrella2019}
C.~Perrella, P.~Light, J.~Anstie, \emph{et~al.}, \enquote{Dichroic two-photon rubidium frequency standard,} {\protect\JournalTitle{Phys. Rev. Appl.}} \textbf{12}, 054063 (2019).

\bibitem{Emily2024}
E.~Ahern, S.~K. Scholten, C.~Locke, \emph{et~al.}, \enquote{Stability characterisation of a two-color two-photon rubidium frequency standard,} {\protect\JournalTitle{In preparation}}  (2024).

\bibitem{Bluebell2024}
S.~K. Scholten, E.~Ahern, C.~Locke, \emph{et~al.}, \enquote{A portable dual-color two-photon rubidium optical frequency standard,} {\protect\JournalTitle{In preparation}}  (2024).

\bibitem{Martin2018}
K.~W. Martin, G.~Phelps, N.~D. Lemke, \emph{et~al.}, \enquote{Compact optical atomic clock based on a two-photon transition in rubidium,} {\protect\JournalTitle{Phys. Rev. Appl.}} \textbf{9}, 014019 (2018).

\bibitem{Lemke2022}
N.~D. Lemke, K.~W. Martin, R.~Beard, \emph{et~al.}, \enquote{Measurement of optical rubidium clock frequency spanning 65 days,} {\protect\JournalTitle{Sensors}} \textbf{22} (2022).

\bibitem{Gray1974}
J.~Gray and D.~Allan, \enquote{A method for estimating the frequency stability of an individual oscillator,} in \emph{28th Annual Symposium on Frequency Control,}  (1974), pp. 243--246.

\bibitem{Allan1981}
D.~Allan and J.~Barnes, \enquote{A modified "allan variance" with increased oscillator characterization ability,} in \emph{Thirty Fifth Annual Frequency Control Symposium,}  (1981), pp. 470--475.

\bibitem{DSAC2021}
E.~A. Burt, J.~D. Prestage, R.~L. Tjoelker, \emph{et~al.}, \enquote{Demonstration of a trapped-ion atomic clock in space,} {\protect\JournalTitle{Nature}} \textbf{595}, 43--47 (2021).

\bibitem{NPLCsFountain}
R.~J. Hendricks, F.~Ozimek, K.~Szymaniec, \emph{et~al.}, \enquote{Cs fountain clocks for commercial realizations—an improved and robust design,} {\protect\JournalTitle{IEEE Transactions on Ultrasonics, Ferroelectrics, and Frequency Control}} \textbf{66}, 624--631 (2019).

\bibitem{RIKEN2020}
M.~Takamoto, I.~Ushijima, N.~Ohmae, \emph{et~al.}, \enquote{Test of general relativity by a pair of transportable optical lattice clocks,} {\protect\JournalTitle{Nature Photonics}} \textbf{14}, 411--415 (2020).

\bibitem{Atkinson2019}
P.~E. Atkinson, J.~S. Schelfhout, and J.~J. McFerran, \enquote{Hyperfine constants and line separations for the $^{1}s_{0}\ensuremath{-}^{3}p_{1}$ intercombination line in neutral ytterbium with sub-doppler resolution,} {\protect\JournalTitle{Phys. Rev. A}} \textbf{100}, 042505 (2019).

\bibitem{sinclair2015invited}
L.~C. Sinclair, J.-D. Desch{\^e}nes, L.~Sonderhouse, \emph{et~al.}, \enquote{Invited article: A compact optically coherent fiber frequency comb,} {\protect\JournalTitle{Review of scientific instruments}} \textbf{86}, 081301 (2015).

\bibitem{Kramer:01}
G.~Kramer and W.~Klische, \enquote{Multi-channel synchronous digital phase recorder,} in \emph{Proceedings of the 2001 IEEE International Frequncy Control Symposium and PDA Exhibition (Cat. No. 01CH37218),}  (IEEE, 2001), pp. 144--151.

\bibitem{Kramer:04}
G.~Kramer and W.~Klische, \enquote{Extra high precision digital phase recorder,} in \emph{2004 18th European Frequency and Time Forum (EFTF 2004),}  (IET, 2004), pp. 595--602.

\end{thebibliography}

\end{document}